\documentstyle[12pt,epsf]{article}
\def\tstrut{\vrule height2.5ex depth0pt width0pt} 
\newcommand{\be}{\begin{equation}}
\newcommand{\bea}{\begin{eqnarray}}
\newcommand{\ee}{\end{equation}}
\newcommand{\eea}{\end{eqnarray}}

\begin{document}
\begin{titlepage}
\begin{flushright}
UG-DFM-1/98 \\
\end{flushright}  
  \vspace*{5mm}

\vspace*{2cm}

\begin{center}
{\Large \bf Bethe-Salpeter Approach  for Meson-Meson 
Scattering in Chiral Perturbation Theory.}

\vspace{1.5cm}
{\large{\bf J. Nieves and E. Ruiz Arriola}}\\[2em]
Departamento de F\'{\i}sica Moderna, Universidad de Granada, 
E-18071 Granada, Spain.

\end{center}

\vspace{2cm}
\begin{abstract}
The Bethe-Salpeter equation restores exact elastic unitarity in the
$s-$ channel by summing up an infinite set of chiral loops. We use this
equation to show how a  chiral expansion can be undertaken
by successive approximations to the potential which should be
iterated. Renormalizability of the amplitudes 
in a broad sense can be achieved  by allowing for 
an infinite set of counter-terms as it is the case in 
ordinary Chiral Perturbation Theory.  
Within this framework we calculate the $\pi \pi$ scattering amplitudes
both for $s-$ and $p-$waves  at lowest order in the proposed expansion
where a successful description of the low-lying resonances ($\sigma$
and $\rho$) and threshold parameters is obtained. We also extract the
$SU(2)$ low energy parameters ${\bar l}_{1,2,3,4}$ from our amplitudes.
\vspace{1cm}

\noindent
{\it PACS: 11.10.St;11.30.Rd; 11.80.Et; 13.75.Lb; 14.40.Cs; 14.40.Aq\\}
\end{abstract}

\end{titlepage}

\newpage

\setcounter{page}{1}

\section{Introduction}

Chiral Perturbation Theory (ChPT) to finite order is unable to
describe resonances. Actually, it is rather the opposite, resonances
determine the bulk of the ${\cal O}(p^4)$ parameters~\cite{GL84,E89}.
Such a description requires the use of a non-perturbative
scheme. Several approaches have been suggested, Pade Re-summation
(PR)~\cite{dht90}, Large $N_f-$Expansion (LNE)~\cite{ei84}, Inverse
Amplitude Method (IAM)~\cite{dp93}, Current Algebra Unitarization
(CAU)~\cite{bbo97}, Dispersion Relations (DP)~\cite{pb97}, Roy
Equations~\cite{W97}, Coupled Channel Lippmann-Schwinger Approach
(CCLS)~\cite{O97} and hybrid approaches~\cite{O98}. Besides their
advantages and success to describe the data in the low-lying resonance
region, any of them has specific drawbacks. In all above approaches
except by LNE and CCLS it is not clear which is the ChPT series of
diagrams which has been summed up. This is not the case for the CCLS
approach, but there a three momentum cut-off is introduced, hence
breaking translational Lorentz invariance and therefore the scattering
amplitude can be only evaluated in the the Center of Mass (CM)
frame. On the other hand, though the LNE and CAU approaches preserve crossing
symmetry, both of them violate unitarity. Likewise, those approaches which
preserve exact unitarity violate crossing symmetry. 

In this paper we propose the use of the Bethe-Salpeter equation (BSE)
to sum up an infinite set of diagrams without use of cut-offs in
the physical amplitudes. We will show that this
re-summation restores exact elastic unitarity in the $s-$ channel and
it naturally leads to the appearance of the experimentally observed
resonances.  Besides, crossing symmetry can be restored perturbatively. Our
approach, at lowest order and in the chiral limit reproduces the
bubble re-summation undertaken in Ref.~\cite{bc93}.

\section{The Bethe-Salpeter Equation.}

\begin{figure}
\vspace{-11cm}
\hbox to\hsize{\hfill\epsfxsize=0.75\hsize
\epsffile[52 35 513 507]{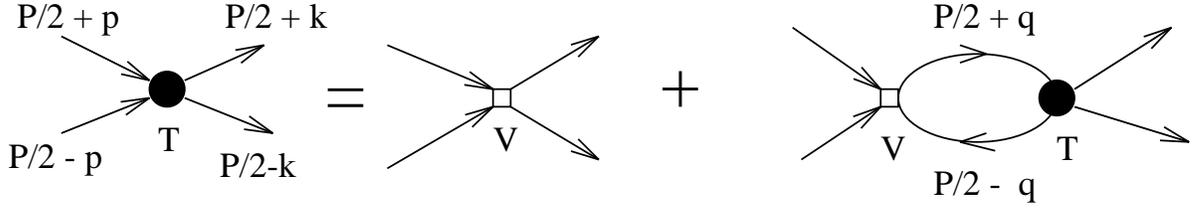}\hfill}
\vspace{1.5cm}
\caption{\footnotesize Diagrammatic representation of the BSE equation. It is also 
sketched the used kinematics.}
\label{fig:kin}
\end{figure}

For the sake of simplicity, we 
neglect coupled channels contributions and thus 
the BSE for the scattering of two identical pseudo-scalar mesons of
mass $m$ and kinematics described in Fig.~\ref{fig:kin}, reads
\begin{eqnarray}
T_P(p,k) &=& V_P(p,k) + {\rm i}\int\frac{d^4
q}{(2\pi)^4}T_P(q,k)\Delta(q+P/2,m) \Delta(-q+P/2,m) V_P(p,q)\label{eq:bs} 
\end{eqnarray}
where $T_P(p,k)$ and $V_P(p,k)$ are the total scattering
amplitude\footnote{The normalization of the amplitude $T$ is
determined by its relation with the differential cross section in the
CM system of the two identical mesons and it is given by $d\sigma
/d\Omega = |T_P(p,k)|^2 / 64\pi^2 s$, where $s=P^2$. The phase of the
amplitude $T$ is such that the optical theorem reads ${\rm Im}
T_P(p,p) = - \sigma_{\rm tot} (s^2-4s\,m^2)^{1/2} $, with
$\sigma_{\rm tot} $ the total cross section.} and the two particle
irreducible amplitude respectively. Besides, $\Delta$ is the exact
pseudoscalar meson propagator. Note that, to solve the above equation
both the off-shell potential and amplitude are required.  Clearly, for
the exact {\it potential} $V$ the BSE provides an exact solution of
the scattering amplitude $T$~\cite{BS51}. Obviously an exact solution
for $T$ is not accessible, since $V$ and $\Delta$ are not exactly
known. We propose an expansion along the lines of ChPT both for the
exact potential ($V$) and the exact propagator ($\Delta$). Thus at
lowest order in this expansion, $V$ should be replaced by the ${\cal
O}(p^2)$ chiral amplitude ($^{(2)}T$) and $\Delta$ by the free meson
propagator, $\Delta^{0}(r,m)=(r^2-m^2+{\rm i}\epsilon)^{-1}$. Even at
lowest order, by solving Eq.~(\ref{eq:bs}) we sum up an infinite set
of diagrams\footnote{Note that our approach is different to the LNE
one~\cite{ei84}. There the bubbles are summed up by means of a
collective field with no well defined isospin. Hence, crossing
symmetry is maintained but elastic unitarity in the $s-$channel is
violated.}. Our expansion is related to the approach recently pursued
for low energy $NN-$scattering where higher order $t-$ and $u-$channel
contributions to the potential are suppressed in the heavy nucleon
mass limit~\cite{k97}.

To illustrate the procedure, let us consider elastic $\pi \pi$
scattering in the $s-$ and $p-$waves. There, for comparison with the 
experimental phase shifts, $\delta_{IJ}(s)$, 
we define the projection over each partial
wave $J$ for each isospin channel $I$
\begin{eqnarray}
T_{IJ}(s) = \frac12 \int_{-1}^{+1}
P_J\left(\cos\theta\right) T_P^I(p,k)~
d(\cos\theta) &=& \frac{{\rm i} 8\pi 
s}{\lambda^{\frac12}(s,m^2,m^2)} 
\left [ e^{2{\rm i}\delta_{IJ}(s)} -1 \right ] 
\end{eqnarray}
where $\theta$ is the angle between $\vec{p}$ and $\vec{k}$, $P_J$
are Legendre polynomials and $\lambda(x,y,z) = x^2+y^2+z^2 -2xy-2xz-2yz$.

\subsection{Isoscalar $s-$wave $\pi \pi$
Scattering. } 
At lowest order, the off-shell potential $V$ in this channel 
is approximated by 
\begin{eqnarray}
V_P^0(p,k) \approx 
^{(2)}T_P^0 (p,k) & = & \frac{5m^2-3s-2(p^2+k^2)}{2f^2}
\end{eqnarray}
where $m$ is the pion mass, for which we take 134.98 MeV, and $f$ the
pion decay constant, for which we take 92.4 MeV.  
To solve Eq.~(\ref{eq:bs}) with the above potential we propose a 
solution of the form
\begin{eqnarray}
T_P^0(p,k) &=& A(s) + B(s) (p^2+k^2) + C(s) p^2 k^2
\end{eqnarray}
where $A,B$ and $C$ are functions to be determined. Note that, as a simple
one loop calculation shows, there appears a new off-shell dependence
($p^2k^2$) not present in the ${\cal O}(p^2)$ potential
$^{(2)}T$. That is similar to what happens in standard ChPT~\cite{GL84}.   

The above ansatz reduces the BSE to a linear algebraic  system of
equations which provides the full off-shell scattering amplitude. The
resulting inverse amplitude on the mass shell and in the CM
frame ($\vec{P}=0, p^0 = k^0 = 0, P^0 = \sqrt{s}$) reads
\begin{eqnarray}
T^{-1}_{00}(s) &=& -I_0(s) +
\frac{2\left(f^2+I_2(4m^2)\right)^2}{2I_4(4m^2)+(m^2-2s)f^2 
+ (s-4m^2)I_2(4m^2) }\label{eq:swave}
\end{eqnarray}
where
\begin{eqnarray}
I_{2n} (s) = {\rm i} \int \frac{d^4
q}{(2\pi)^4}\frac{(q^2)^n}
{\left[(q-\frac{P}{2})^2-m^2+{\rm i}\epsilon\right]
\left[(q+\frac{P}{2})^2-m^2+{\rm i}\epsilon\right]}
\end{eqnarray}
$I_0, I_2$ and $I_4$ are ultraviolet divergent integrals and thus
Eq.~(\ref{eq:swave}) requires renormalization. We have made use of
translational and Lorentz invariance which relate the integrals
$I_2(s)$ and $I_4(s)$ with $I_0(s)$ and the divergent constants
$I_2(4m^2)$ and $I_4(4m^2)$. Note also that $I_0(s)$ is only
logarithmically divergent and it only requires one subtraction, ie.,
${\bar I}_0 (s) = I_0(s)-I_0(4m^2)$ is finite and it is given by  
\begin{eqnarray}
{\bar I}_0 (s) &=& \frac{1}{(4\pi)^2} \sqrt{1-\frac{4m^2}{s}}  \log
\frac{\sqrt{1-\frac{4m^2}{s}}+1 }{\sqrt{1-\frac{4m^2}{s}}-1}
\end{eqnarray}
where the complex phase of the argument of the $\log$ is taken in the
interval $[-\pi,\pi[$

To renormalize the amplitude given in Eq.~(\ref{eq:swave}), we note
that in the spirit of an Effective Field Theory (EFT) all possible
counter-terms should be considered. This can be achieved in our case
in a perturbative manner, making use of the formal expansion of the
bare amplitude $T = V + VG_0V + VG_0VG_0V+ \cdots $, where $G_0$ is
the two particle propagator. Thus, a counter-term series should be
added to the bare amplitude such that the sum of both becomes
finite. At each order in the perturbative expansion, the divergent
part of the counter-term series is completely determined. However, the
finite piece remains arbitrary as long as the used {\it potential} $V$
and the pion propagator are approximated rather than being the exact
ones. Our renormalization scheme is such that the renormalized
amplitude can be cast, again, as in Eq.~(\ref{eq:swave}).  This
amounts in practice, to interpret the previously divergent quantities
$I_{0,2,4}(4m^2)$ as renormalized free parameters. After having
renormalized, we add a superscript $R$ to differentiate between the
previously divergent, $I_{0,2,4}(4m^2)$, and now finite quantities,
$I^R_{0,2,4}(4m^2)$. These parameters and therefore the renormalized
amplitude can be expressed in terms of physical (measurable)
magnitudes.  In principle, these quantities should be understood in
terms of the underlying QCD dynamics, but in practice it seems more
convenient so far to fit $I^R_{0,2,4}(4m^2)$ to the available data.
The threshold properties of the amplitude (scattering length,
effective range, etc..) can then be determined from them.  Besides the
pion properties $m$ and $f$, at this order in the expansion we have
three parameters.  The appearance of three new parameters is not
surprising because the highest divergence we find is quartic
($I_4(s)$) and therefore to make it convergent we need to perform
three subtractions. This situation is similar to what happens in
standard ChPT where one needs to include at the one loop level some
new low-energy parameters (${\bar l}_i$)~\cite{GL84}. In fact, if $t-$
and $u-$ channel unitarity corrections are Taylor expanded around
threshold, a comparison of our (now) finite amplitude,
Eq.~(\ref{eq:swave}), to the $ {\cal O} (p^4) \,\, \pi\pi$ amplitude
in terms of these parameters, ${\bar l}_{1,2,3,4}$, becomes possible. 
We do this explicitly in the Appendix. 

At the lowest order in the expansion proposed in this work, we
approximate, in the scattering region $s>4m^2$, the ${\cal O}(1/f^4)$
$t-$ and $u-$ channel unitarity corrections (function $h_{IJ}$ in
Eq.~(\ref{eq:gl})) by a Taylor expansion around threshold to order
$(s-4m^2)^2$. At next order in our expansion (when the full ${\cal
O}(p^4)$-corrections are included both in the {\it potential} and in
the pion propagator) we will recover the full $t-$ and $u-$ channel
unitarity logs at ${\cal O}(1/f^4)$, and at the next order (${\cal
O}(1/f^6)$) we will be approximating these logs by a Taylor expansion
to order $(s-4m^2)^3$. Thus the analytical structure of the amplitude
derived from the left hand cut is only recovered perturbatively. This
is in common to other approaches (PR, DP, IAM $\cdots$) fullfiling
exact unitarity in the s-channel, as discussed 
in~\cite{dp93},~\cite{pb97},\cite{GM91}.

\subsection{Isovector $p-$wave $\pi \pi$ Scattering} 

At lowest order, the off-shell potential $V$ in this channel 
is approximated by 
\begin{eqnarray}
V_P^1(p,k) \approx 
^{(2)}T_P^1(p,k) & = & \frac{2p \cdot k}{f^2}
\end{eqnarray}
As before, to solve Eq.~(\ref{eq:bs}) with the above potential we propose a 
solution of the form
\begin{eqnarray}
T_P^1(p,k) &=& M(s) p\cdot k + N(s)  (p\cdot P)( k\cdot P) 
\end{eqnarray}
where $M$ and $N$ are functions to be determined. Note that, as
expected from our previous discussion for the $s-$wave, 
there appears a new off-shell dependence
($(p\cdot P)( k\cdot P) $) not present in the ${\cal O}(p^2)$ potential.
Again, this ansatz reduces the BSE to a linear algebraic  system of
equations which provides the full off-shell scattering
amplitude. The resulting inverse amplitude on the mass shell, after angular
momentum projection,  and in CM frame reads
\begin{eqnarray}
T^{-1}_{11}(s) =  &=&
-I_0(s) + \frac{ 2I_2(4m^2)-6f^2}{s-4m^2}\label{eq:pwave}
\end{eqnarray}
Similarly to the $s-$wave case, the above equation presents
divergences which need to be consistently removed in terms, for
instance, of the scattering volume and effective range in the
$p$-wave. Because the highest divergence present is quadratic only two
subtractions are needed and hence  only two undetermined parameters
appear. As we will discuss below, 
crossing symmetry at ${\cal O}(1/f^4)$ provides a relationship between
the parameters entering in different isospin channels.
We emphasize once more, that the proliferation of undetermined
parameters is not a drawback as compared to standard ChPT.

\begin{figure}
\vspace{-1.5cm}
\begin{center}                                                                
\leavevmode
\epsfysize = 550pt
\hspace{-1cm}\makebox[0cm]{\epsfbox{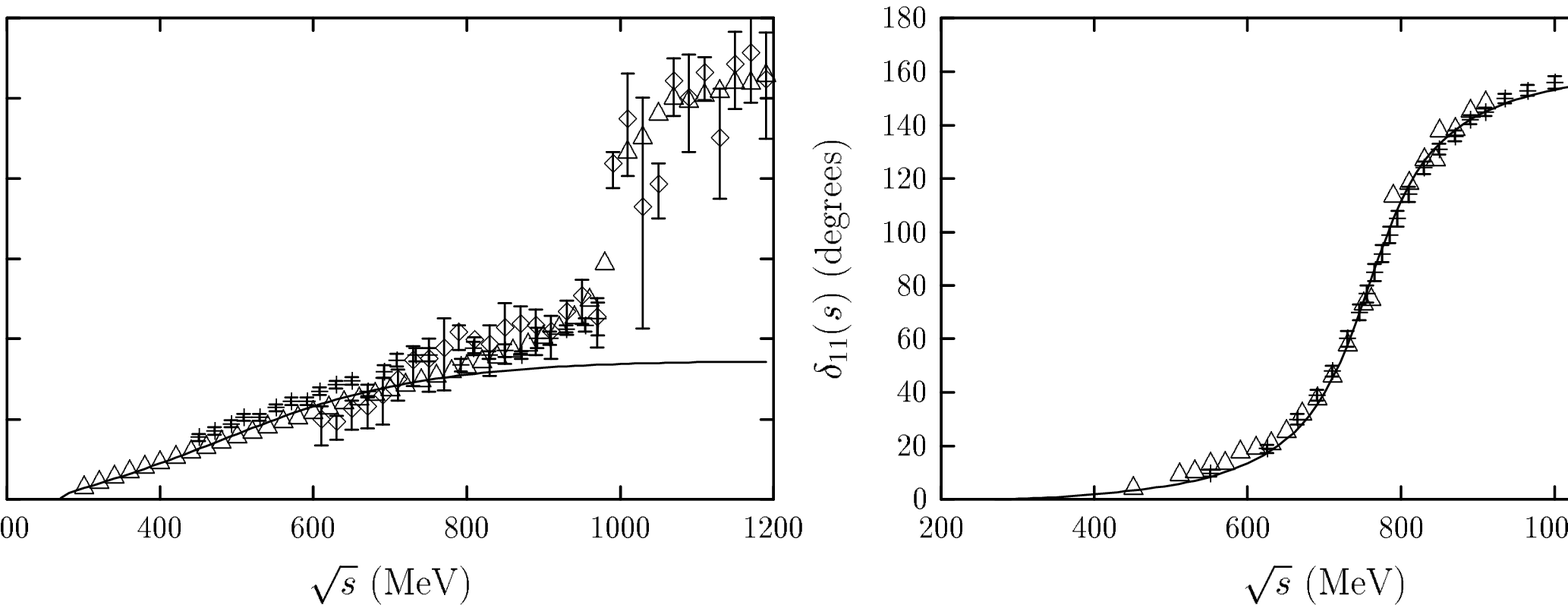}}
\end{center}
\vspace{-12.5cm}
\caption{\footnotesize $s-$ (left) and $p-$wave (right) $\pi\pi$ 
phase shifts as a function of the total CM
 energy $\protect\sqrt s$. Left panel: Triangles, squares and
crosses stand for the experimental analysis of 
Refs.~\protect\cite{fp77},~\protect\cite{klr97} 
and~\protect\cite{e73}, respectively. 
Right panel: Triangles and
crosses stand for the experimental analysis of 
Refs.~\protect\cite{em79} and~\protect\cite{pa73}, respectively.  Solid lines 
are the fit (see Table~\protect\ref{tab:spwaves}) of our $s$-channel unitarized
${\cal O}(p^2)-$model to the data of Refs.~\protect\cite{fp77} 
and~\protect\cite{pa73}.}
\label{fig:spwaves}
\end{figure}

\section{Results}

The solutions of Eqs.~(\ref{eq:swave}) and (\ref{eq:pwave})  
satisfy elastic unitarity. Indeed, 
the imaginary part of $T^{-1}_{IJ}(s)$ is determined by 
the imaginary part of ${\bar I}_0 (s)$ and in the physical region is given by
\begin{eqnarray}
{\rm Im }T^{-1}_{IJ}(s+{\rm i}\epsilon) 
= - {\rm Im }{\bar I}_0 (s+{\rm i}\epsilon) &=& 
 \frac{\lambda^{\frac12}(s,m^2,m^2)}{16\pi s}
\end{eqnarray}
in agreement with the unitarity requirement provided $m$ is  the 
physical pion mass. That 
makes possible to extract the phase shifts unambiguously within our
framework.

As we have already mentioned, our approach violates crossing
symmetry. At order ${\cal O}(1/f^4)$ our
isoscalar $s-$ and isovector $p-$wave amplitudes are polynomials
of degree two in the variable $(s-4m^2)$, with a total of five (3+2)
arbitrary coefficients (see Eq.~(\ref{eq:ej})), and there are no
logarithmical corrections to account for $t-$ and $u-$channel
unitarity corrections ($h_{IJ}$ term in Eq.~(\ref{eq:gl})). Far from
the left hand cut, these latter corrections can be expanded in a
Taylor series to order $(s-4m^2)^2$, but in that case the one loop
$SU(2)$ ChPT $s-$ and isovector $p-$wave amplitudes can be cast as
second order polynomials in the variable $(s-4m^2)$, with a total of
 four (${\bar l}_{1,2,3,4}$) arbitrary coefficients~\cite{GL84}. 
To restore, in this approximation, crossing symmetry in our amplitudes 
requires the existence of a relationship between our five undetermined
parameters. This relation reads (see the Appendix)
\begin{equation}
75 I^{R,s}_2/2m^2 + 8I^{R,p}_0+
33I^{R,s}_0+5I^{R,p}_2/m^2+\frac{10157}{1920\pi^2} = 0 \label{eq:cons}
\end{equation}
Once this constraint is implemented in our model, there exists a linear
relation between our remaining four undetermined parameters and the
most commonly used ${\bar l}_{1,2,3,4}$ parameters (Eq.~(\ref{eq:li})).

In Fig.~\ref{fig:spwaves} we show the agreement of our model with the
experimental phase shifts, both in the $s-$ (left) and
$p-$(right)waves. We fit the four undetermined parameters\footnote{We use
Eq.~(\protect\ref{eq:cons}) to express $I^{R,s}_2$ in terms of
$I^{R,p}_{0,2}$ and $I^{R,s}_0$.}  $I^{R,s}_{0,4}(4m^2)$ and
$I^{R,p}_{0,2}(4m^2)$ to the scalar and vector data. Results of the
combined fit can be found in Table~\ref{tab:spwaves}. Values of
$\chi^2/$dof and the threshold parameters deduced from our formulae
are also shown in Table~\ref{tab:spwaves}. As we see, the vector
channel is satisfactorily described up to 1 GeV, whereas the scalar
channel is well reproduced up to 0.8 GeV. In the latter case, and for
these high energies, one should also include the mixing with the
$K\bar K$ channel as pointed out recently in
Refs.~\cite{O97}--\cite{O98}. Regarding the deduced threshold
parameters we find agreement within experimental uncertainties. For
completeness we quote the deduced values for the low energy
coefficients (see Eq.~(\ref{eq:li}) in the Appendix)
\begin{equation}
{\bar l}_1 = -0.90 \pm 0.09,\,\,{\bar l}_2 = 6.03 \pm 0.08,\,\,
{\bar l}_3 = 2 \pm 7,\,\, {\bar l}_4 = 3.9 \pm 0.4 \label{eq:eles}
\end{equation}
if a fit to the data of Refs.~\protect\cite{fp77},~\protect\cite{em79}
is performed\footnote{We remind to the reader that we have assumed a
5\% error in the data of these two references. If we had assumed a
1(10)\% errors, the statistical fluctuations
 quoted in Eq.~(\protect\ref{eq:eles}) would have
decreased (increased) roughly by a factor of 5 (2).}.  As we see they are
in reasonable good agreement with those obtained in standard ChPT
\cite{GL84}, which also use the same set of data to analyze $\pi\pi$
scattering. When the data of Ref.\cite{pa73} for the $p-$wave are
fitted we find compatible values for ${\bar l}_{1,2}$ and incompatible
values for ${\bar l}_{3,4}$. We prefer to use the data of
Ref.~\protect\cite{em79} to estimate the low energy parameters ${\bar
l}_{1,2,3,4}$, since they come much closer to threshold than those of
Ref.~\protect\cite{pa73} ($\sqrt s = 350 $ MeV versus 450
MeV). Furthermore both sets of data are incompatible up to about
$\sqrt s = 650 $ MeV.

Once we have determined the 
parameters ${\bar l}_{1,2,3,4}$, the $I=2,J=0$  lowest order BSE
amplitude is completely fixed, as demanded from crossing symmetry.  It
is to say, the corresponding integrals which appear in this channel
are completely determined by the ${\bar l}\,'s$ or equivalently by the
$\sigma -$ and $\rho -$channel integrals. We have evaluated the
corresponding phase-shift in the $I=2,J=0$ channel, using the values
of Eq.~(\ref{eq:eles}), and compared to the data of Ref.~\cite{ho77}
finding a remarkable agreement up to about $1450$ MeV ($\chi^2 /{\rm
(num.\, data)} = 0.9$) and threshold
parameters ($m\,a_{20}=-0.0390\pm 0.0018$, and $m^3\,b_{20}=-0.0701\pm
0.0010$)  in agreement within uncertainties with the
experimental values.
\begin{table}
\vspace{-0.3cm}
\begin{center}
\begin{tabular}{c|cc|cc}
& \multicolumn{2}{c|}{Data
Refs.~\protect\cite{fp77},~\protect\cite{pa73}} & 
\multicolumn{2}{c}{Data Refs.~\protect\cite{fp77},~\protect\cite{em79}}\\
& $J=I=0$ &  $J=I=1$&  $J=I=0$ &  $J=I=1$\\\hline\tstrut
$I^R_0(4m^2)$ & $-0.0323 \pm 0.0005 $ & $-0.1157 \pm 0.0017$
& $-0.0289 \pm 0.0005 $ & $-0.0980 \pm 0.0017$\\\tstrut
$I^R_2(4m^2)$ & $-$ & $0.011\pm 0.006  $ 
& $-$ & $0.072\pm 0.005  $ \\\tstrut
$I^R_4(4m^2)$ & $-0.031 \pm 0.004  $ & $-$
& $-0.003 \pm 0.003  $ & $-$\\\hline\tstrut 
$m^{2J+1} a_{IJ}$ & $0.233 \pm 0.011 $ & $0.0292 \pm 0.0005$ 
& $0.204 \pm 0.008 $ & $0.0356 \pm 0.0006$ \\\tstrut
$m^{2J+3} b_{IJ}$ & $0.237 \pm 0.007 $ & $0.0050 \pm 0.0002$
& $0.267 \pm 0.006 $ & $0.0062 \pm 0.0002$ \\\hline
$\chi^2/$dof & \multicolumn{2}{c|}{$1.6$} 
& \multicolumn{2}{c}{$3.9$} \\\tstrut
$I^{R,s}_2(4m^2)$  & \multicolumn{2}{c|}{$0.0082 \pm 0.0008$}
& \multicolumn{2}{c}{$-0.0016 \pm 0.0007$} \\\hline
\end{tabular}
\end{center}
\caption{\footnotesize Fitted ($I^R_n(4m^2)$) parameters to the
experimental data of Refs.~\protect\cite{fp77}
($J=I=0$),~\protect\cite{em79} and~\protect\cite{pa73} ($J=I=1$).
$I_n(4m^2)$ are given in units of $(2m)^n$ and $I^{R,s}_2(4m^2)$ is
calculated from Eq.~(\protect\ref{eq:cons}).  Errors in the fitted
parameters are statistical and have been obtained by increasing the
value of $\chi^2$ by one unit.  We also give the threshold parameters
$a_{IJ}$ and $b_{IJ}$ obtained from an expansion of 
the scattering amplitude, Re$T_{IJ}
= -16\pi m (s/4 -m^2)^J [ a_{IJ} + b_{IJ} (s/4 -m^2) + \cdots ]$ close
to threshold. In the the $s-$wave channel and due to the
lack of error estimates in Ref.~\protect\cite{fp77}, we have assumed a
rule of thumb error of 5\% in the data and carry out the fit up to
900 MeV. We have chosen this set of data because the data of
Refs.~\protect\cite{klr97} and~\protect\cite{e73} seem to be
inconsistent between each other at low energies and cover near
threshold a narrower region than that covered by the analysis of
Ref.~\protect\cite{fp77}. For the $p-$wave, the data of
Refs.~\protect\cite{em79} (here again, we assume an error of 5\% in
the data, because the lack of error estimates of the original analysis
) and ~\protect\cite{pa73} disagree again, specially close to
threshold. }
\label{tab:spwaves}
\end{table}

Comparison to recent two-loop
calculations~\cite{mksf95} would also be possible. To
make such a comparison really meaningful, we should include in our
analysis higher waves than the $s-$ and $p-$waves considered here,
which would necessarily require to include ${\cal O}(p^4)$ corrections
to both the {\it potential} and the pion propagator. This point is out of the
scope of this work and will be subject of future
research~\cite{ej99}. Nevertheless, we would like to point out that
the lowest order explored in this letter, though very simple, it
already leads to reasonable sizes for the two loop contributions to
the amplitudes. Thus for instance, for the $s-$wave scattering length we
obtain $a_{00}/a_{00}^{\rm tree} = 1 + 0.27\,({\rm one\,\, loop})\, +
0.07 \,({\rm two \,\, loops}) + \cdots$, in a reasonable agreement
with the findings of Bijnens {\it et al.,} in Ref.~\cite{mksf95}.

Recent work, Refs.~\cite{O97}--\cite{O98}, suggests that off-shellness
can be ignored when solving the BSE, since it only amounts to a pion
mass and pion decay constant renormalization.  This on-shell procedure
corresponds to setting our $I^R_2(4m^2)$ and $I^R_4(4m^2)$ parameters
to zero, and taking $f$ and $m$ as the physical parameters. Direct
inspection of our expressions shows that one can not simultaneously
absorb the $\sigma-$ and $\rho-$channel divergences by a redefinition
of the $m$ and $f$ parameters. Since the model of
Refs.~\cite{O97}--\cite{O98} has been constrained not to have these
new parameters ($I^R_{2,4}(4m^2)$) generated by the off-shellness it
is impossible for them to simultaneously reproduce the $\sigma-$ and
$\rho-$channels; a polynomial, with some new parameters, has to be
added to the on shell-constrained solution of the BSE to properly
describe the resonance region in the $J=I=1$ channel. The inclusion of
this polynomial with free parameters is
justified~\cite{O97}--\cite{O98} within the IAM
approach~\cite{dp93}. We would like to stress here that considering
explicitly the off-shellness automatically embodies the IAM, as it is
invoked in Refs.~\cite{O97} and \cite{O98}, and thus incorporates such
a polynomial.

\section{Conclusions and Outlook.} 

We have solved the BSE for $\pi\pi$ scattering in the ladder
approximation. This is to say using the {\it potential} and pion
propagator at lowest order in the chiral expansion. This calculation
produces unitary amplitudes in the $s-$channel and describes
satisfactorily $s-$ and $p-$ wave phase-shifts from threshold up to
the region of low-lying resonances.  The present approach can be
extended in principle to higher orders in the chiral expansion, i.e.,
including in the {\it potential} and in the pion propagator terms of
order ${\cal O}(p^4)$ and higher.  The new divergences will become
more severe and more subtraction constants will be needed. That will
be translated in an increasing number of free parameters, as it is the
case also of standard ChPT. To be more specific, for instance, the
${\cal O}(p^4)$ {\it potential} contains unitarity corrections in the
$t-$ and $u-$ channels, (the corresponding $s-$channel correction to
the amplitude is not two particle irreducible). Hence, the solution of
the BSE with this {\it potential} will automatically implement exact
unitarity in the $s-$channel and perturbative unitarity in the $t-$
and $u-$channels.

The inclusion of unitarity corrections in the $t-$ and $u-$ channels
in the {\it potential} makes the practical solution of
the BSE a cumbersome task and it is currently underway~\cite{ej99}. However,
if one neglects these corrections in the $t-$ and $u-$ channels the BSE
can be solved using a much simpler algebraic procedure as shown here for the
lowest order case. In that case only unitarity in the $s-$channel will
be restored and crossing symmetry will be violated. Thus, one should expect a 
general solution, below the four pion production threshold, of the form 
\begin{eqnarray}
T^{-1}_{IJ}(s)  = -{\bar I}_0(s) +
\frac{P^{IJ}_{n}(s-4m^2)}{Q^{IJ}_n(s-4m^2)}
\label{eq:pade}
\end{eqnarray}
where $P^{IJ}_n(x)$ and $Q^{IJ}_n(x)$ are polynomials\footnote{Note
that the first non-vanishing coefficient in $Q_n$  corresponds to the 
power $(s-4m^2)^J$.} of order $n$
(corresponding to the order ${\cal O}(p^{2n})$ in the proposed expansion)
with real coefficients. Most of these coefficients have to be fitted
to the data to accomplish with the renormalization program. 
The resemblance to a sort of {\it pade}
approximant for the inverse amplitude, although not
exactly in the form proposed in Ref.~\cite{dht90}, is striking. 
On the other hand,
Eq.~(\ref{eq:pade}) provides a model of the type $N/D$~\cite{CM60} for the
amplitude, where $D$ has a right cut and the function $N$ is approximated
by a polynomial with no cuts. These ideas have been recently examined
in Ref.~\cite{oo98}. 

Finally, we should mention that to describe the $s-$wave at higher
energies than 0.8 GeV, a coupled channel formalism needs to be
used. Such an improvement of our model can be easily implemented, 
at least at the lowest order ${\cal O}(p^2)$.

\section*{Acknowledgments}
We would like to acknowledge useful discussions with J.A. Oller and
E. Oset. This research was supported by DGES under contract PB95-1204 and by
the Junta de Andaluc\'\i a.

\section*{ Appendix}

The   $ {\cal O}(p^2)+{\cal O}(p^4)$ on-shell $SU(2)-$ ChPT 
isoscalar $s-$ and isovector 
$p-$ wave $\pi \pi$ amplitudes, expressed in terms of the 
renormalization  invariant parameters ${\bar l}_{1,2,3,4}$, 
are given by~\cite{GL84}
\begin{eqnarray}
T_{IJ}(s)&=&  V^{IJ}_1(s) / f^2 + \left ( V^{IJ}_2(s)+  {\bar I}_0
(s)\times [\,V^{IJ}_1(s)\,]^2\right )  / f^4 \label{eq:gl}\\\nonumber
&&\\\nonumber
V^{IJ}_1(s)&=&\left\{\begin{array}{ll}\frac{m^2-2s}{2}&I=J=0 \\
&\\
\frac{4m^2-s}{6}&I=J=1\end{array}\right. \\\nonumber
&&\\\nonumber
V^{IJ}_2(s)&=&-\frac{1}{192\pi^2}g_{IJ}+ 
\frac{1}{12}h_{IJ} \\\nonumber
&&\\\nonumber
h_{IJ}&=& \int_{-1}^1 \frac{d(\cos\theta )}{2}\left ( 
f_{IJ}(t){\bar I}_0 (t) + f_{IJ}(u){\bar I}_0 (u)\right )
P_J(\cos\theta )\\\nonumber
&&\\\nonumber
&=& \left\{\begin{array}{ll} \frac{5m^4}{4\pi^2}+ \frac{101 m^2
(s-4m^2)}{96\pi^2} + \frac{191 m^2 (s-4m^2)^2}{288\pi^2}+ \cdots
&I=J=0 \\ &\\ \frac{89 m^2
(s-4m^2)}{288\pi^2} - \frac{37 m^2 (s-4m^2)^2}{2880\pi^2}+ \cdots
&I=J=1\end{array}\right.
\end{eqnarray}
where $t= -2(s/4-m^2)(1-\cos\theta )$, $u= -2(s/4-m^2)(1+\cos\theta
)$ and
\begin{eqnarray}
f_{IJ}(x)&=&\left\{\begin{array}{ll} 10x^2 + x(2s-32m^2)+37m^4-8sm^2
&I=J=0 \\ 2x^2 + x(s+2m^2)-m^4-4sm^2
&I=J=1 \end{array}\right. \\\nonumber
&\\\nonumber
g_{IJ}&=&\left\{\begin{array}{ll} m^4\left(40{\bar l}_1+ 80{\bar l}_2-
15{\bar l}_3+84{\bar l}_4+125\right) + m^2(s-4m^2)\times &\\
\left(32{\bar l}_1+ 48{\bar l}_2+24{\bar l}_4+\frac{232}{3}\right)+  
(s-4m^2)^2\left(\frac{22}{3}{\bar l}_1+ \frac{28}{3}{\bar l}_2+
\frac{142}{9}\right)&I=J=0 \\ & \\ 
\frac{s-4m^2}{3}\left [ 4m^2 \left(-2{\bar l}_1+2{\bar l}_2
+3{\bar l}_4+1 \right)+ (s-4m^2)\left(-2{\bar l}_1+2{\bar l}_2+1 
\right)\right]&I=J=1\end{array}\right. 
\end{eqnarray}

On the other hand, expanding up to order ${\cal O}(p^4)$ the
amplitudes of Eqs.~(\ref{eq:swave}) and (\ref{eq:pwave}) we find,
\begin{eqnarray}
T_{IJ}(s) & = & V^{IJ}_1(s) / f^2 + \left ( W^{IJ}(s)+  {\bar I}_0
(s)\times [\,V^{IJ}_1(s)\,]^2\right )  / f^4 \label{eq:ej}\\\nonumber
&&\\\nonumber
W^{IJ}(s)&=&\left\{\begin{array}{ll} I^{R,s}_4+ 7m^2I^{R,s}_2
+  \frac52 (s-4m^2)
I^{R,s}_2 + I^{R,s}_0\times [\,V^{00}_1(s)\,]^2 &I=J=0 \\ & \\ 
V^{11}_1(s) 
\left (\frac13I^{R,p}_2+I^{R,p}_0\,V^{11}_1(s)\right) 
&I=J=1 \end{array}\right.
\end{eqnarray} 
Taylor expanding around threshold the function $h_{IJ}$ and
identifying the functions $V^{IJ}_2$ of Eq.~(\ref{eq:gl}) and 
$W^{IJ}_2$ of Eq.~(\ref{eq:ej}) one obtains the constraint given in 
Eq.~(\ref{eq:cons}) and the following relations:
\begin{eqnarray}
{\bar l}_1 &=& \frac{107}{750} + \pi^2 \left (\frac{112}{25}I^{R,p}_0-
\frac{288}{25}I^{R,s}_0\right )\label{eq:li}\\ 
&&\nonumber\\
{\bar l}_2 &=& -\frac{1997}{3000} -\pi^2 \left (\frac{352}{100}I^{R,p}_0+\frac{1152}{100}I^{R,s}_0\right) \nonumber\\ 
&&\nonumber\\
{\bar l}_3 &=&
 -\frac{2761}{3750} +\pi^2 \left (\frac{1472}{375}I^{R,p}_0-
\frac{1776}{125}I^{R,s}_0+ \frac{224}{75m^2}I^{R,p}_2 
+ \frac{64}{5m^4}I^{R,s}_4\right) \nonumber\\ 
&&\nonumber\\
{\bar l}_4 &=& \frac{173}{120} + \pi^2 \left (
\frac{16}{3}I^{R,p}_0+
\frac{8}{3m^2}I^{R,p}_2\right)\nonumber
\end{eqnarray}

\end{document}